\documentclass{mn2e}

\usepackage{graphicx}

\title[QPO Spectra in the Lense--Thirring precession model]
{Energy spectra of X-ray quasi-periodic oscillations in the Lense--Thirring
precession model}

\newcommand\rt{R_{\rm tr}}
\newcommand\rms{r_{\rm ms}}
\newcommand\Rdisc{r_{\rm disc}}
\newcommand\lh{l_{\rm h}}
\newcommand\ls{l_{\rm s}}
\newcommand\lszero{l_{\rm s,0}}
\newcommand\thetap{\theta_{\rm p}}
\newcommand\phip{\phi_{\rm p}}
\newcommand\MSun{M_{\odot}}
\newcommand\taur{\tau_{\rm r}}

\author[P. T. \.{Z}ycki, C. Done, A. Ingram]
{Piotr T. \.{Z}ycki$^{1}$\thanks{e-mail: ptz@camk.edu.pl}, Chris Done$^2$, 
Adam Ingram$^3$ \\
    $^1$Nicolaus Copernicus Astronomical Center, Polish Academy of Sciences, 
       Bartycka 18, 00-716 Warsaw, Poland \\
    $^2$Centre for Extragalactic Astronomy, Department of Physics, University of Durham, South Road, Durham DH1 3LE\\
    $^3$ Anton Pannekoek Institute, University of Amsterdam, Science Park 904, 1098 XH Amsterdam, The Netherlands}

\date{Accepted ... Received ...; in original form ...}
\pubyear{2016}

\begin{document}
\label{firstpage}
\pagerange{\pageref{firstpage}--\pageref{lastpage}}

\maketitle

\begin{abstract}

We model the energy dependence of a quasi periodic oscillation (QPOs) produced by 
Lense-Thirring precession of a hot inner flow. We use a fully 3-dimensional 
Monte-Carlo code to compute the Compton scattered flux produced by the hot inner 
flow intercepting seed photons from an outer truncated standard disc. The changing 
orientation of the precessing torus relative to the line of sight produces 
the observed modulation of the X-ray flux. 

We consider two scenarios of precession. First, we assume that the precession
axis is perpendicular to the plane of the outer disc. In this scenario
the relative geometry of the cold disc and the hot torus does not change 
during precession, so the emitted spectrum does not change, and the modulation is
solely due to the changing viewing angle.
In the second scenario the precession axis is tilted with respect to the 
outer disc plane. This leads to changes in the relative geometry of 
the hot flow and cold plasma, possibly resulting in variations of the plasma
temperature and thus generating additional spectral variability, which combines with the
variations due to the viewing angle changes. 

\end{abstract}


\section{Introduction}

\begin{figure*}
  \includegraphics[width=2\columnwidth]{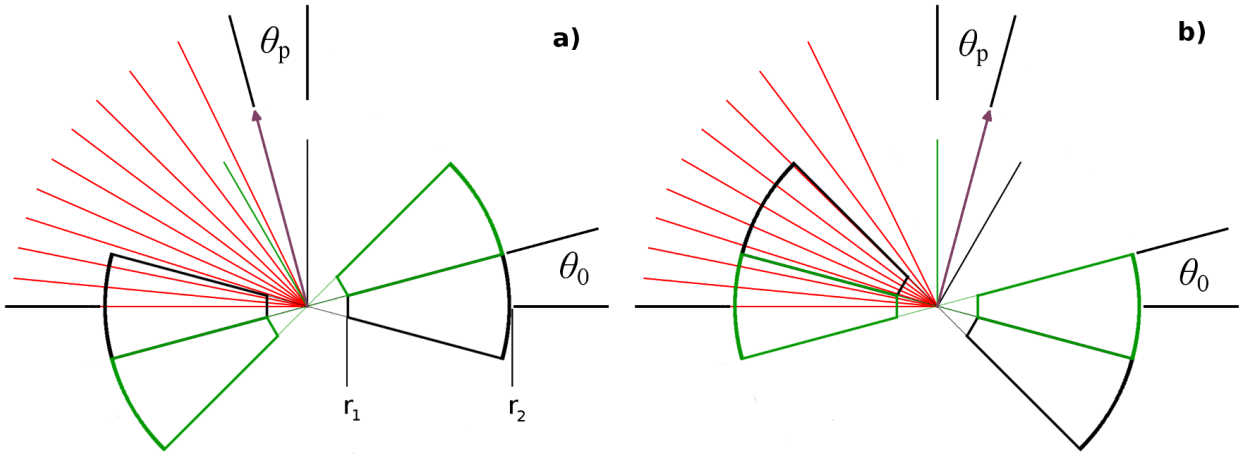}
 \caption{
Schematic view of the geometry in the scenario 2). The precession axis (with arrow; brown) 
is tilted to the outer disc normal by angle $\theta_p$.
The two panels show two extreme orientations of the precession axis relative to the
direction towards the observer, located to the left, in the plane of the page.
Each panel shows in turn two extreme positions of the precessing torus (green and black)
relative to the observer. 
The viewing angle is represented by the red fan-like lines, spaced by $\cos i = 0.1$. 
The torus thickness is measured by its half-opening angle, $\theta_0$, as marked. 
For the assumed $\theta_0 = 15^\circ$ and $\theta_P=15^\circ$, the observer usually sees 
the top of the torus in case a), while they would see more of the outer torus boundary in case b).
See also fig~2 in Veledina et al.\ (2013).
\label{fig:geom}}
\end{figure*}

The low-frequency QPO is a common feature observed in the 1-10 Hz range of 
X-ray power spectra of standard (i.e., $\sim 10\,\MSun$) black hole binaries.
Their frequencies vary and are correlated with the slope of the X-ray spectrum,
suggesting a connection of the QPO with a variable truncation radius of the standard 
accretion disk, as it makes a transition into a hot inner flow.
Moreover, the QPO energy spectra (or the rms$(E)$ amplitude) imply that it is 
the harder, Comptonized component of the energy spectrum that 
undergoes the oscillations, rather than the directly observed disc emission, although
the driving parameter of the variability may depend on the spectral state of the source 
(e.g., Churazov, Gilfanov \& Revnivtsev 2001; \.{Z}ycki \& Sobolewska 2005; 
Sobolewska \& \.{Z}ycki 2006; Remillard \& McClintock 2006; 
Done, Gierli\'{n}ski \& Kubota 2007; 
Axelsson, Done \& Hjalmarsdotter 2014, Axelsson \& Done 2016).

One possible mechanism to explain the low-$f$ QPO is the Lense-Thirring precession
of the inner disc, appearing as a result, for example, of the accreting plasma
forming a disc in a plane inclined to the equatorial plane of a rotating black hole
(Bardeen-Petterson effect; Bardeen \& Petterson 1975; Fragile, Mathews \& Wilson 2001).
Lense-Thirring precession was suggested as a model of QPO in X-ray binaries
by Stella \& Vietri (1998). That initial model, based on a simple solution of the precession
predicted, among other things, that the frequency of precession is a strong
function of the black hole spin. This is actually a problem because the observed
frequencies of low-$f$ QPO show rather little scatter, not neceesarily 
correlated with the estimated spin values of black holes in relevant sources 
(e.g. Pottschmidt et al.\ 2003).

For geometrically thick torii, simulations show that the misalignment results in 
a solid body precession of the inner flow, rather than the flow forming a disk 
in the equatorial plane of the black hole (Fragile et al.\ 2007). 
Moreover, the inner edge of the torus appears to be cutoff by bending waves
in such a way that the precession frequency becomes only weakly dependent on
the black hole spin (Ingram, Done \& Fragile 2009). 
These two recent results make the Lense-Thirring precession
not only an attractive but also a realistic model to explain the 
low-$f$ QPO, especially as the same 
geometry of an inner hot flow and outer cold disc
is the best scenario 
to explain the low/hard state of accreting sources.
Specific models formulated within this geometry are able to
explain basic spectral and variability properties as well as correlations between 
spectral and timing parameters observed in X-ray binaries (Done et al.\ 2007).

\begin{figure}

  \includegraphics[width=\columnwidth]{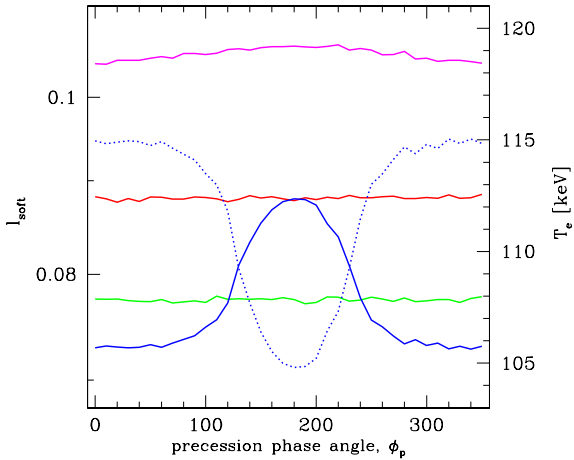}

 \caption{Ratio of the disc luminosity intercepted by the torus to the total 
disc emitted luminosity, $l_{\rm soft}=\ls/\lszero$ (solid curves; left y-axis), as a function
of precession phase, $\phip$. 
The red curve shows a reference 
case of the torus co-planar with the outer disc. The green curve shows our precession 
scenario 1), where the torus precesses around an axis perpendicular to the outer
disc plane, so the relative geometry of the two does not change. However, the torus tilt 
means that the fraction $\ls/\lszero$ is lower than in the reference case. (Note that in these 
two situation $\ls/\lszero$ is constant -- the fluctuations demonstrate the statistical 
noise in the Monte Carlo simulations.)
The two blue curves show the precession scenario 2), where the changing relative geometry
leads to the $\ls/\lszero$ ratio varying between the maximum value corresponding to 
the co-planar 
configuration (at $\phip=180^{\circ}$) and the minimum value, 
corresponding to tilt $2\thetap$. The red, green and blue curves are for the
torus opening $\theta_0 = 15^\circ$, while the magenta curve shows the case
of $\theta_0= 45^\circ$ (geometrically thicker torus) in the scenario 2) (so it should
be compared with the blue curve). The dotted blue curve shows the variations of the
electron temperature, corresponding to the blue solid curve of $\ls/\lszero$.
\label{fig:lhls}}
\end{figure}

The overall geometry which we thus envision would be that of a standard
Shakura-Sunyaev accretion disc truncated at a given radius, $\rt > \rms$, where
$\rms$ is the marginally stable circular orbit. The inner accretion flow,
below $\rt$, consists of the hot plasma forming precessing torus. 
Hard X-ray are produced by inverse Compton upscattering of the soft photons 
entering the torus from the cold disc.

The mechanisms of modulation of X-ray emission in this geometry are two-fold.
Firstly, the angle between the line of sight and the axis of the torus changes
with precession phase. Since the Comptonized emission in this geometry is not 
isotropic, the observed emission varies. Secondly, plasma temperature
changes, if the relative geometry of the hot and cold phases of accretion changes,
because of the change of heating to cooling ratio of the plasma. This leads to
spectral (slope) variations and adds to the former variability.
A third effect, variable relativistic distortions of the radiation, may also be 
important. Although the above mentioned truncation of the inner radius 
means that the flow does not really extend into the region of very strong gravity, 
the light bending of the photons emitted on the opposite side of the torus introduces
distortions to the observed light curves. 
This will be especially important for high inclination sources.

The model based on the above ideas has been explored in a number of papers.
Ingram \& Done (2011, 2012a) showed how the precession scenario fits into the global model
of X-ray spectra and variability of accreting black holes. It does indeed fit well with
the standard explanation of the low/hard state, namely the truncated disc/inner hot flow
geometry, and the propagating fluctuations model explaining the broad band X-ray 
variability. 

Ingram \& Done (2012b) added to the model the Fe K$\alpha$ line, 
produced in the outer disc as a result of X-ray illumination by the inner flow. 
They showed that the varying illumination of the outer disc by the precessing flow
produces a characteristic modulation of the centroid energy of the line with the QPO
period.
The recent detection of this signal in H 1743-322 gives strong support to the model 
(Ingram et al.\ 2016a).

Veledina \& Poutanen (2015) 
generalized the effect of X-ray reprocessing in the outer disc to compute the
optical continuum component from the reprocessing and they predicted how variability of 
the X-ray and optical flux may be correlated in this model.
Finally, Ingram et al.\ (2015)
computed the polarisation of the X-ray radiation and showed how this changes as a result
of the precession.

One uncertain aspect of the Lense-Thirring precession model is the excitation 
mechanism of the precession. Presumably, if the precession were caused by the
misalignment of the orbital plane and equatorial plane of the black hole,
it would be a persistent phenomenon. Observationally, the low-$f$ QPO are transitory
phenomena and their appearance seems to correlate with state transitions of the
sources. This would suggest that the precession would have to be somehow induced 
internally
within the flow as its structure changes, but it remains to be seen if such an
effect is indeed possible. One possible difference in precession geometry between
the two possibilities might be the orientation of the precession axis. It would
be tilted to the outer disc axis, in the misalignment case, while it might be 
expected to be aligned with the outer disc axis, if the precession is somehow
induced internally. A hint about the geometry may be coming from recent observations
of H 1743-322, where Ingram et al.\ (2016b) find variations of the amplitude of the 
reflected component, pointing out to the misalignment case (our scenario 2, see below).

The goal of this paper is to determine the character of spectral variability
expected in this model in a realistic 3-D geometry. We perform Monte Carlo simulations 
of the inverse Compton
upscattering of the soft photons from the outer disc in the hot uniform precessing torus, 
taking into account all the usual
physical processes and the geometrical setup. We simulate sequences of energy 
spectra from the precession, for a range of torus parameters, study the spectral
variability and compute the variability characteristics, for example, the amplitude 
as a function of energy,
rms$(E)$. We  consider the two above mentioned cases of the orientation of 
the precession axis: perpendicular to the outer disc plane (scenario 1) and 
tilted with respect to it (scenario 2). We do not include the reflection component
resulting from reprocessing of the hard X-rays in the outer disc. This will be included
in the forthcoming investigation. We do not include the broad band variability of 
the comptonized emission, either, so the presented rms$(E)$ refers to the QPO only, but
it does include all the Fourier harmonics.

\section{Model}

\subsection{Geometry}

\begin{figure*}

  \includegraphics[width=2\columnwidth]{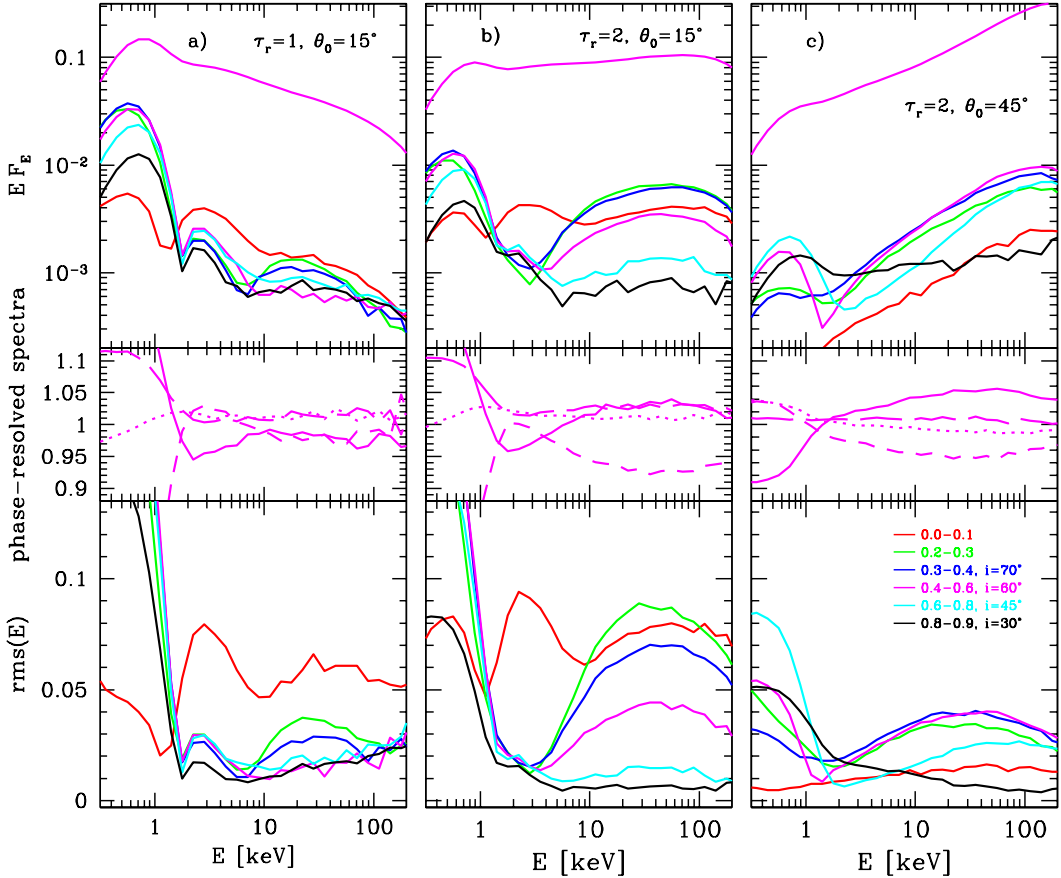}

 \caption{Compton scattered spectra and QPO variability for geometry 1) (precession axis 
perpendicular to the plane of the outer disc).
Columns a), b), c) show the results for different torus parameters:
radial optical thickness, $\taur$, and the half-opening angle $\theta_0$, as labelled. 
The top panels in each column show the time averaged spectrum at the viewing angle 
$i=60^\circ$ (topmost, magenta curve) and the QPO spectra for different viewing angle, as labelled 
(description in bottom panel, column c). 
The middle panels show the QPO-phase resolved spectra at four phases separated by $90^\circ$, normalized to the time averaged spectrum, at $i=60^\circ$, to demonstrate the range of 
spectral variability.
The bottom panels show the fractional ${\rm rms}(E)$ amplitude for different viewing angle.
See Sec~\ref{sec:prec1rms} for details.
\label{fig:prec1rms}}
\end{figure*}

We assume that the hot inner flow forms a torus with a wedge-like cross-section 
(Fig.~\ref{fig:geom}), with the half-opening angle, $\theta_0$,
measured from the equatorial plane. The inner boundary, at $r_1$, is assumed 
cylindrical, while the outer boundary, at $r_2$, is spherical. 
The outer cold disc extends from $\Rdisc\approx r_2$ outwards. 
The hot torus precesses around a precession axis, with (constant) precession angle 
(polar angle between the axis of the torus and the precession axis) 
$\theta_p$. As discussed in Introduction the precession axis itself can be 
1) perpendicular to the plane
of the outer disc, or 2) it can be inclined to the plane by 
the same angle $\theta_p$.  The major difference between
the two geometries is that in scenario 1) the relative geometry of the hot 
torus and the cold disc does not change with the precession
phase, while it does change in geometry 2). In fact, in scenario 2),
at one particular phase
of precession, the equatorial planes of the hot flow and the cold disc may overlap,
so the former is not inclined to the latter. 
There is one more parameter in scenario 2), which is the azimuthal position angle of 
the precession axis, $\phi_0$. 
We use a convension that $\phi_0 = 0^\circ$ means that the precession axis is directed 
towards the observer (Fig.~\ref{fig:geom}a) while
$\phi_0 = 180^\circ$ means that the precession axis is directed away from the observer 
(Fig.~\ref{fig:geom}b). 
Schematic representation of the geometry can also be seen in Ingram \& Done (2012b;
fig~1) and in fig~2 in 
Veledina et al.\ (2013), where the angle  $\beta$ corresponds to our $\theta_p$.

\begin{figure*}

  \includegraphics[width=2\columnwidth]{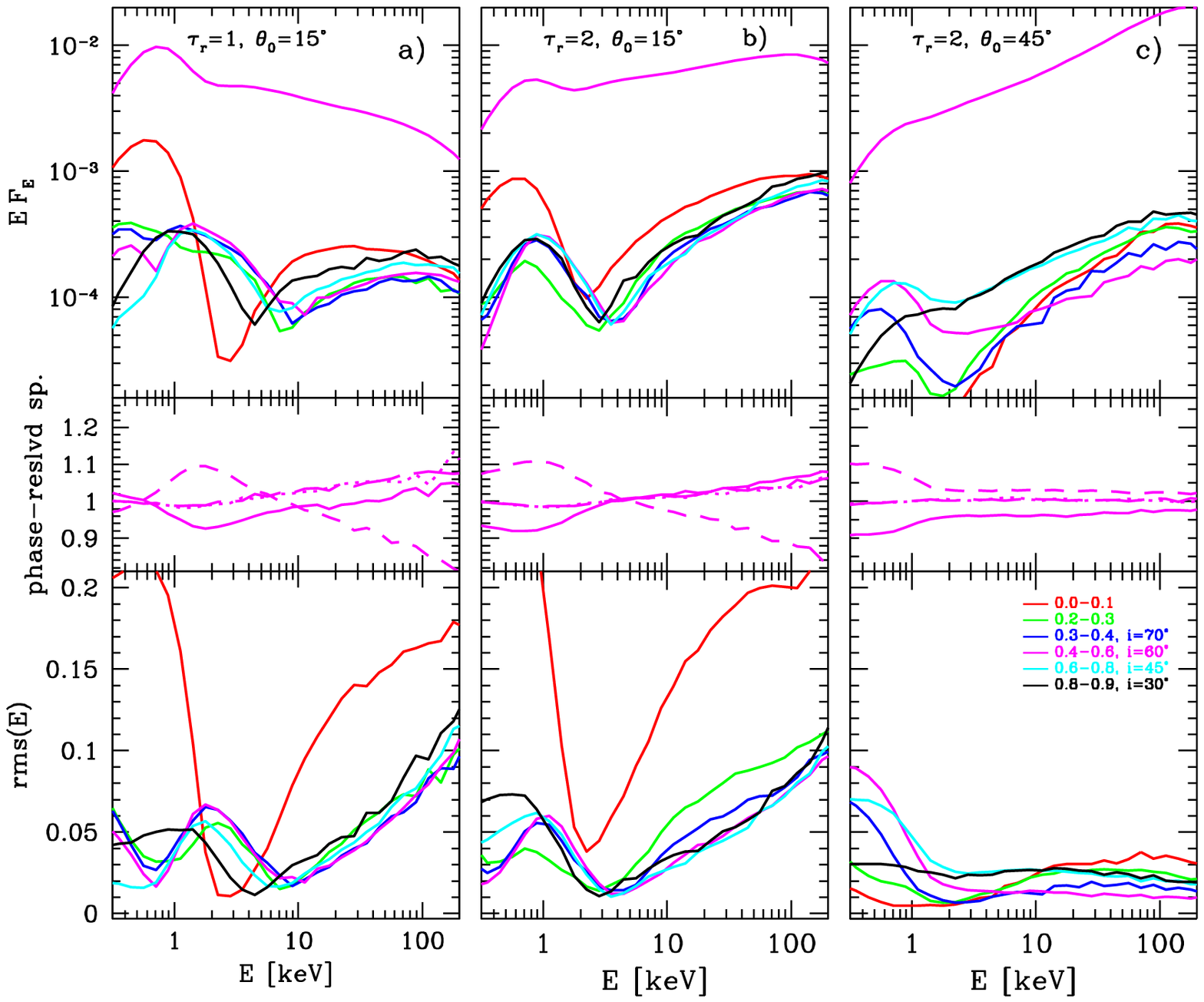}

\caption{Compton scattered spectra and QPO variability for geometry 2), for the case of 
$\phi_0=0^\circ$ (precession axis inclined wrt to the outer disc normal, directed towards
the observer; see fig~\ref{fig:geom}a). 
Columns a), b), c) show the results for different torus parameters: $\taur$ and 
$\theta_0$, as labelled. 
The top panels in each column show the time averaged spectrum at the viewing angle $i=60^\circ$ 
(topmost, magenta curve) 
and the QPO spectra for different viewing angle, as labelled in the bottom panel c). 
The middle panels show the QPO-phase resolved spectra at four phases separated by $90^\circ$, 
normalized to the time averaged spectrum, at $i=60^\circ$. 
The bottom panels show the ${\rm rms}(E)$ amplitude for different viewing angle. 
\label{fig:prec2rms0}}
\end{figure*}

\begin{figure*}

  \includegraphics[width=2\columnwidth]{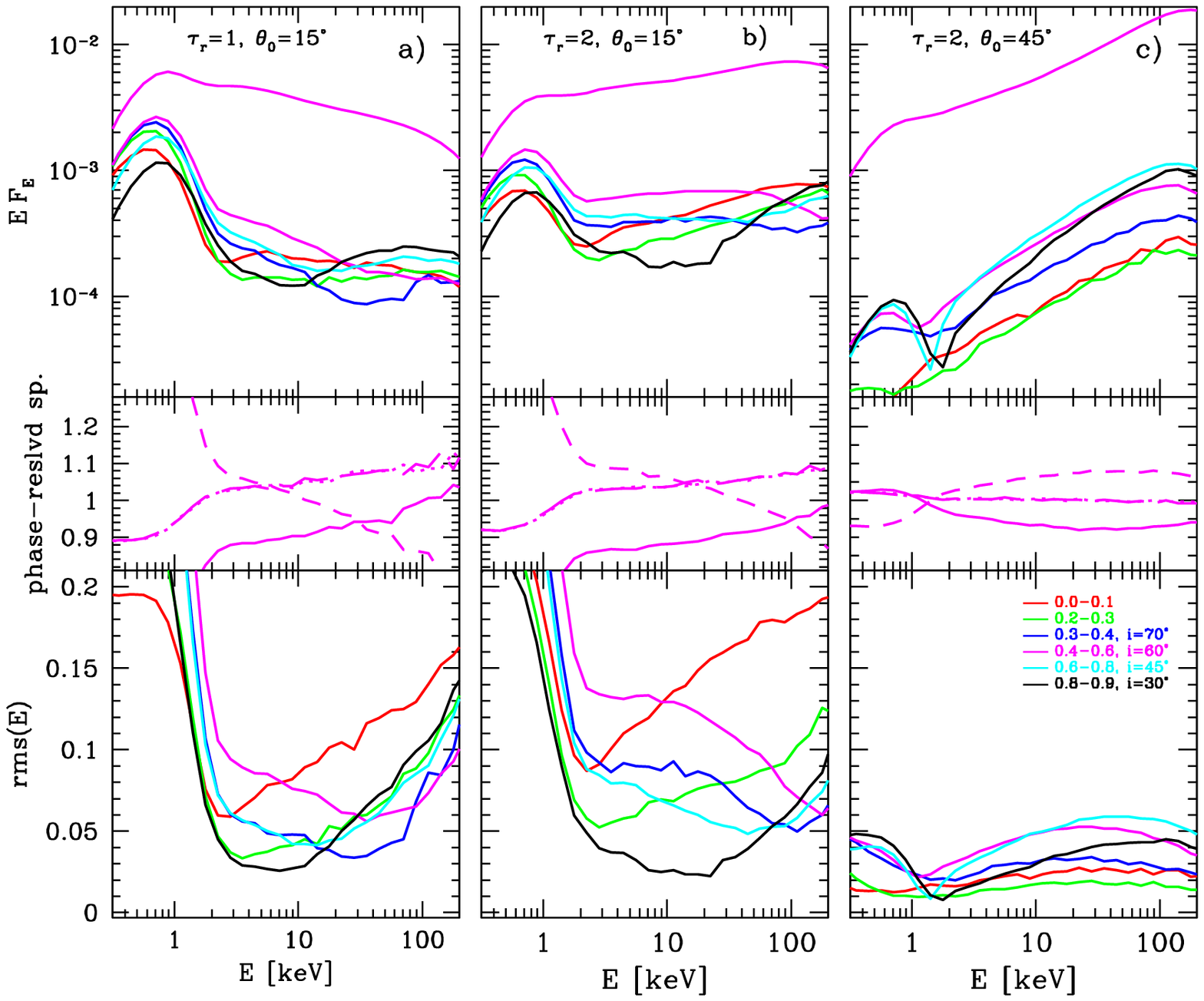}

 \caption{Compton scattered spectra and QPO variability for geometry 2), for the case of 
$\phi_0=180^\circ$ (precession axis inclined wrt to the outer disc normal, directed away from
the observer; see fig~\ref{fig:geom}b). 
Columns a), b), c) show the results for different torus parameters: $\taur$ and 
$\theta_0$, as labelled. 
The top panels in each column show the time averaged spectrum at the viewing angle $i=60^\circ$ 
(topmost, magenta curve) 
and the QPO spectra for different viewing angle, as labelled in the bottom panel c). 
The middle panels show the QPO-phase resolved spectra at four phases separated by $90^\circ$, 
normalized to the time averaged spectrum, at $i=60^\circ$. 
The bottom panels show the ${\rm rms}(E)$ amplitude for different viewing angle. 
\label{fig:prec2rms2}}
\end{figure*}

The optical thickness of the hot plasma in the radial direction is $\tau_r$,
the thickness in other directions following from the assumed uniform density
of the plasma, and corresponding dimensions. 
The cold disc is assumed to radiate a local blackbody emission, with
the standard radial temperature profile $T(r) \propto r^{-3/4}$, 
i.e.\ we neglect any boundary effects (e.g., illumination). Consistently,
we neglect the hard X-ray component reflected/reprocessed by the cold disc. 
The local disc emission is assumed isotropic. 
The maximum disc temperature, at $r=\Rdisc$, is denoted $T_0$.

\subsection{Monte Carlo comptonization code}

The code was already described in Janiuk, Czerny \& \.{Z}ycki (2000). 
It uses the standard prescriptions of Pozdnyakov, Sobol \& Syunyaev (1993) 
and G\'{o}recki \& Wilczewski (1984) for simulating the inverse Compton
scattering of soft photons in a hot plasma. We implemented a range of 
geometries, allowing the soft photons to originate both inside the hot
plasma and outside it (as is the present case). The code assumes uniform
plasma density and temperature and it does not include the X-ray reflection from
the outer disc. In the present case we consider full 3-D
geometry, schematically shown in Fig.~\ref{fig:geom}.

\section{Results}

The inner hot flow is quite likely to be geometrically thick, so we consider
the cases of $\theta_0=15^\circ$ ($h/r \approx 0.3$) and 
$\theta_0=45^\circ$ ($h/r = 1$). The absolute (physical) values of the 
inner and outer torus radii would be important for determination
of the plasma temperature (through the heating-cooling balance), but
with our simplified treatment of the temperature we use dimentionless units.
Nevertheless, the ratio $r_2/r_1$ is important and we assume $r_1 = 6$ and $r_2=30$.
The precession angle is assumed $\theta_p = 15^\circ$, while $T_0 = 0.2\,$keV.
We compute a sequence of spectra corresponding to different phases of
precession and calculate the r.m.s.\ amplitude of variability as a function of photon
energy  in the usual way.

\subsection{Plasma temperature}

The hot torus intercepts a fraction of the soft photons from the cold 
disc. In the geometrical scenario 1) this fraction is constant as a function
of precession phase, while it does vary in geometry 2). 
This is demonstrated in Fig.~\ref{fig:lhls}, where we plot the ratio of 
luminosity intercepted by the torus to the total luminosity emitted by the disc, 
for two values of $\theta_0$.

The intercepted luminosity is $\approx 10\%$, for all considered parameters.
In the scenario 2), and for the geometrically thick torus, $\theta_0 = 45^\circ$, 
the variations are very small, $\approx 2\%$. This is because, for
$\theta_p < \theta_0$ (i.e., the torus is geometrically thicker than the range of
its wobble due to precession), only the outer torus boundary is visible from the 
outer disc, irrespective of the precession phase. Thus, the effective solid angle
of the torus from the outer disc is approximately constant, considering
that the disc emission is concentrated towards its inner edge. 
Variations are somewhat larger, $\approx 20\%$, for the thinner torus, 
when the effective
solid angle of the torus does depend on the degree of misalignement of the 
torus and disc planes. 

The varying fraction of the intercepted soft flux implies 
variations of the plasma temperature, $kT$, since it affects the heating 
to cooling ratio of the plasma, $\lh/\ls$.
Physical values of $\lh$ and $\ls$ would depend on the details of energy
generation in the flows, most importantly the radial energy generation rates
and the transition radius, $r_2 = \Rdisc$. Detailed considerations of
these processes are clearly beyond the scope of this paper, therefore here
we simply assume a constant value of $\lh$, and $\ls$ varying as described 
above.

We estimate variations of $kT$ adopting Comptonization formulae from 
Beloborodov (1999a), with $\tau_T=1$ and $\delta=1/6$. 
The average $\lh/\ls$ ratio is assumed such that the
resulting plasma temperature is close to 100 keV, as observed in
the accreting sources.
The temperature is constant in geometry 1) as the seed photons are constant with phase, but
the change in $l_{\rm s}$ for geometry 2) means that $kT(\phi)$ also changes by around 
10\%, as shown in Fig.~\ref{fig:lhls} (dotted line).

\subsection{Variability amplitude}

Here we present the fractional amplitude of variability (containing both the main
QPO and its all harmonics) as a function of energy, in the two considered scenarios 
for the precession geometry.
We want to compare our simulations with the observations which show subtle changes in 
QPO shape relative to the Compton scattered continuum (Axelsson et al.\ 2014; 
Axelsson \& Done 2016). Hence we plot only the Compton scattered spectra and its 
variability,  and do not include the constant disc component which will dilute the 
fractional variability of both the QPO and broadband variability at low energies. 

We consider two values of the radial optical thickness of the torus,
$\tau_r = 1, 2$ and two values of the torus shape, parameterized by $\theta_0$,
with $\theta_0 = 15^\circ$ and $45^\circ$. We combine these torus parameters into three cases,
$\tau_r = 1$, $\theta_0 = 15^\circ$; $\tau_r = 2$, $\theta_0 = 15^\circ$ and
 $\tau_r = 2$, $\theta_0 = 45^\circ$, resulting in a broad range of the ``effective'' 
optical thickness of the torus, which determins the slope of the final Comptonized spectrum.

\subsubsection{Geometry I. Precession axis perpendicular to the outer disk}
\label{sec:prec1rms}

The results are plotted in Fig.~\ref{fig:prec1rms}. Three columns of the plot show the
three cases of the ``effective'' optical thickness. The top panels of each column present
the time average spectrum  at the viewing angle $i=60^\circ$ (the top magenta curve) 
and the QPO spectra (i.e., the variable components), for different viewing angle 
as labelled (viewing angle labels in the bottom panel, column c). 

The middle panels show the QPO-phase resolved spectra, 
at four values of the QPO phase separated by $90^\circ$, 
normalized to the time averaged spectrum, at $i=60^\circ$, to demonstrate the range of 
spectral variability.

The bottom panels show the ${\rm rms}(E)$ amplitude for different viewing angle.

The highest
amplitude of the Comptonized component alone is actually observed at energies 
corresponding to seed photon spectrum. Related to this, there is a pronounced bump 
in rms$(E)$, especially for high viewing angle for thin torus ($\thetap = 15^\circ$),
at energies corresponding to the first scattering in the Comptonized spectrum. 
As the torus orientation changes relative to the observer,
the effective optical thickness along the line of sight changes. This results
in a variable fraction of scattered and unscattered photons.
The change of effective $\tau$ (``effective'' meaning integrated over all
lines of sight leading to different points on the emitting disc) depends on
the viewing angle. For high viewing angle (close to ``edge on'') the change is large,
since depending on the torus position, an observer may see through the entire torus, 
or very little of it. For lower viewing angle (close to ``top view'') there is still
a large amplitude of variability of the un-scattered component, because of the 
possibility of (part of) the torus to be in the line of sight at a particular
precession phase. However, the observed 
scattered component will have reached observer after changing its direction of motion
by about $90^\circ$. Photons originating at {\em any\/} azimuthal angle are 
approximately equally likely to achieve this, so such a component will only be weakly
variable with the precession phase.
Obviously, the observed amplitude at energies of the seed photons is strongly 
suppressed when the directly observed, constant disc emission is added
(Fig.~\ref{fig:prec1rms}).

The feature in rms$(E)$ due to first scattering is very prominent for low $\tau$
and/or high viewing angle situations. However, its strength might be a consequence of 
the simplifying
assumptions of the model, most importantly the uniform temperature and density of the
plasma. The situation might be reminiscent of that from early studies of
Comptonized spectra (e.g., Stern et al.\ 1995). These predicted strong first-scattering 
feature in the spectrum, which was not observed in X-ray data. It was later realized
that a realistic geometry would be rather non-uniform and that would most
likely smear out such features.

Where the first-scattering feature is not so prominent, the rms$(E)$ increases 
somewhat
with energy (at least for viewing angles $\ge 30^\circ$), 
up to a maximum at 20--40 keV, then it decreases again. This behaviour
reflects the subtle changes of the spectrum, as the changing orientation of the torus
means it is observed at changing viewing angle. 
The variability is stronger for larger $\tau_r$ (right vs.\ left panels
in Fig.~\ref{fig:prec1rms}), at least
when the line of sight passes through the torus, because the transmitted fraction,
$e^{-\tau}$, is more variable. At low viewing angles the variability is generally weaker,
since the emission is closer to isotropy. The energy dependence is also weaker,
the rms spectra being either independent of energy or slighly deceasing with $E$.

Variability is weaker for geometrically thicker torus (right vs.\ left panels in 
Fig.~\ref{fig:prec1rms}). At high viewing angle the reason for this is that now
the optical depth along the line of sight does not change much, as the range
of torus ``wobble'' (due to precession) is small compared to torus angular size.
At low viewing angle, the reason is rather different, namely it is the higher degree
of isotropy of emission for thicker (geometrically hence also optically) torus.

\subsubsection{Geometry II. Precession axis tilted to the outer disk}
\label{sec:prec2rms}

In this scenario there is one additional parameter, which is the location of
the observer relative to the precession axis. 
The location can be parameterized by the position
angle of the precession axis, $\phi_0$, i.e.\ the 
angle between the plane of the precession axis and outer disc normal, and 
the plane of the line of sight and the outer disc normal (angle $\phi$ in fig~2 of 
Veledina et al.\ 2013; see Fig~\ref{fig:geom}). 
For $\phi_0$ close to $0^\circ$ the observer (especially positioned close to top-view)
may see only the top of the torus and never see 
its outer boundary, while for $\phi_0$ close to $180^\circ$  the observer may
see the outer boundary for a significant fraction of precession period. 
Obviously, situations with $\phi_0$ and $360^\circ-\phi_0$ are equivalent.
Figures~\ref{fig:prec2rms0}--\ref{fig:prec2rms2} show the resulting rms$(E)$
dependencies for different $\phi_0$.

In this geometry the amplitude of variability can be expected to be larger than in 
geometry I.
There are two obvious reasons for this, both related to the variable number of soft
photons entering the hot precessing torus. The mere fact that the numer of soft photons
varies causes additional variability of the normalization of the overall Comptonized
spectrum. Then there is the additional spectral variability caused by temperature 
variations of the plasma. However, the obvious correlation between these two parameters
(Fig.~\ref{fig:lhls}), and the complicated dependence of the spectrum on the precession 
angle (Sec.~\ref{sec:prec1rms}) mean that the resulting rms$(E)$ is rather complicated.
In particular, the influence of the temperature variations is rather different for 
$\phi_0 = 0^\circ$ and for $\phi_0 = 180^\circ$.
The total variability amplitude is 4-10\% for our parameters which approximately matches 
the observed range of values.

The results show a complicated dependence of rms$(E)$ on the viewing angle.
 The dependence is actually rather weak when mostly top/bottom torus walls
are observed, as is the case of $\phi_0 = 0$ at not-too-high viewing angle 
(Fig~\ref{fig:prec2rms0}a,b), or when the torus emits almost isotropically 
because of larger optical depth (Fig~\ref{fig:prec2rms0}c; Fig~\ref{fig:prec2rms2}c). 
However, there is a significant
dependence on $i$ when during the torus precession an observer sees both the outer
and upper torus walls (Fig~\ref{fig:prec2rms2}a,b).

\section{Discussion}
\label{sec:discuss}

The Lense-Thirring precession appears to be an attractive model for the low-frequency
QPO observed in black hole binaries. Current magnetohydrodynamical simulations reveal
that a hot torus-shaped plasma can  precess as a solid body
(Fragile et al.\ 2007), with frequencies
matching those observed in black hole X-ray binaries (Ingram et al.\ 2009). 
This precessing hot torus scenario fits well into the overall geometry envisioned for
the hard state of X-ray binaries, where a trucated disk and a hot inner flow seems 
to be the most complete scenario, explaining, at least at the phenomenological level, 
a range of observational data both in the spectral and time domain 
(review, e.g., in Done et al.\ 2007). The recent detection of the predicted iron line 
energy modulation with QPO phase (Ingram et al. 2016a) gives strong support to this model.

The most important variability characteristic presented in this paper is the 
energy dependence of the amplitude of variability. The model fractional rms$(E)$ does not 
generally show a clear monotonic dependence. 
There are various effects influencing the rms$(E)$, like the seed photon contribution and 
the first scattering effects. Also, if the precession results in the observer seeing both
the top and the outer torus boundary, the rms$(E)$ seems to be more complicated than in the
simpler case (cf Fig.~\ref{fig:prec2rms0} and Fig.~\ref{fig:prec2rms2} and panels a) and b) 
in Fig.~\ref{fig:geom}), although in a realistic situation the hot plasma may form a rather
smoother structure, which may influence the details of rms$(E)$.
There are nevertheless regions of parameter space where rms$(E)$ does increase, which can also
be represented as the QPO energy spectrum being harder than the time averaged spectrum.
The character of the simulated rms$(E)$ relation can be expected to be only weakly dependent
on the spectral shape, as the mechanism of the modulation is independent of the
spectral formation. 

Observationally, an interesting anticorrelation seems to exist between 
the spectral shape (slope) of the time averaged and the QPO spectrum:
the QPO spectrum is harder than the time averaged spectrum in the soft spectral state
of a source, while the QPO spectrum is softer than the time averaged spectrum when
the source is in a hard specral state  (e.g., Sobolewska \& \.{Z}ycki 2006; Axelsson et al.\ 2014;
Axelsson \& Done 2016). 
Simple analysis performed by \.{Z}ycki \& Sobolewska (2005) suggests that the two types
of behaviour correspond to two different driving parameters of the oscillations, 
for example, either by the plasma heating rate or the cooling rate. A particular geometrical
scenario considered in  \.{Z}ycki \& Sobolewska (2005), namely a modulation of the covering 
factor of the cold reflecting plasma (e.g.\ the cold disc), predicted the QPO spectrum being 
softer than the time averaged spectrum. 

The Lense-Thirring precession model is a different 
geometrical model, though, and for some parameters it does predict the QPO spectrum harder 
than the time averaged specrum, as mentioned above. This happens for a range of torus
shapes, producing both hard and soft time averaged spectra. It is therefore uncertain that
it would be possible to construct such a series of models that would on one hand predict spectral
hard--soft state evolution and, consistently the QPO spectrum evolution.

The model predicts also a significant variability of the seed photons component.
This is because of the varying observed fraction of the disc component 
transmitted through the precessing torus. The observed soft emission would obviously be a sum
of the transmitted fraction and the directly observed disc emission and the latter dominates
over the former (the intercepted fraction is about 10\%, Fig.~\ref{fig:geom}) and so the
observed variability amplitude would be negligible at the seed photon energy, as indeed observed.

\begin{figure}

  \includegraphics[width=\columnwidth]{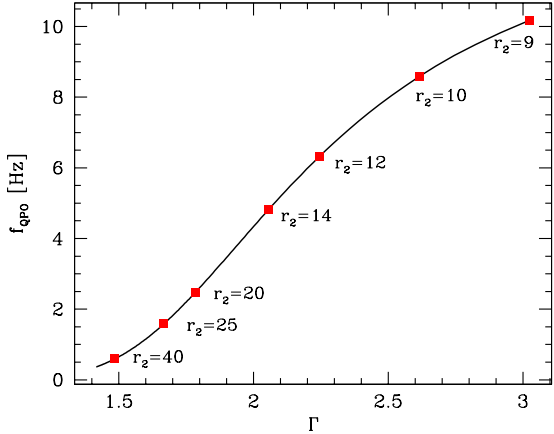}

 \caption{Relation between the spectral slope of the Comptonized emission, $\Gamma$,
and the QPO frequency, as a function of the transition radius between the outer disc
and the precessing torus, $r_2$, for black hole spin $a=0.9$. The spectral slope depends on 
$r_2$ through the plasma heating-to-cooling ratio, while $f_{\rm QPO}$ depends on $r_2$ because
the precessing torus moment of inertia depends on $r_2$ (Ingram et al.\ 2009). 
See Discussion for details.
\label{fig:GmFr}}
\end{figure}

The presented Lense-Thirring precession model can also qualitatively reproduce the correlation
between the spectral slope and the QPO-frequency. 
Adopting a simple description of the energy generation in the accretion flow,
\begin{equation}
 F_{\rm grav} (r) \propto \left(1-\sqrt{r_1 \over r}\right) {1 \over r^3},
\label{equ:fg}
\end{equation}
where the inner radius of the precessing torus is given as 
$r_1 = 3.0 (h/r)^{-4/5} a^{2/5}$ (Ingram et al.\ 2009  and references therein), 
we can describe the heating-cooling balance of the hot plasma as a function of the transition 
radius, $r_2$:
\begin{equation}
 l_{\rm h}(r_2) \propto \int_{r_1}^{r_2} r F_{\rm grav}(r) dr,
\end{equation}
and  $l_{\rm s}(r_2)= 0.1 [l_{\rm tot}-l_{\rm h}(r_2)]$, where 
$l_{\rm tot} = \int_{{\rm r}_1}^{\infty} r F_{\rm grav}(r)\,dr$, and the numerical factor 0.1 
accounts for the fraction of soft photons intercepted 
by the hot flow (Fig.~\ref{fig:lhls}). Using $\Gamma=2.33(l_{\rm h}/l_{\rm s})^{-1/6}$ 
(Beloborodov 1999b) and combining it with Eq.~2 in Ingram et al.\ (2009) we can obtain
a simple example of the $\Gamma$-$f_{\rm QPO}$ relation, for a given
value of $a$ (Fig.~\ref{fig:GmFr}).
Obviously, Eq.~\ref{equ:fg} is a rather {\em ad hoc\/} generalization of the simple
formula for gravitational energy dissipation in Keplerian disc in Schwarzschild
metric, so it should not be relied upon to provide a robust quantitative relation. However, 
it does capture the trend and the right range of both parameters.

The presented computations are admittedly simplified in some aspects. Most importantly
the Comptonizing plasma is uniform in density and temperature, and the situation
described is quasi-static. In more realistic scenarios the overall variability would come
from perturbations of mass accretion rate propagating radially through the flow.
A perturbation in mass flow may result in a perturbation of plasma temperature
propagating inwards, leading to spectral evolution superposed on the geometrical
effect. This may remove from the spectra the signatures of the first scattering, which are generally
not observed (e.g., Stern et al.\ 1995).
These more complicated scenarios should be considered in  detail to see
whether they may account for the whole range of observed behaviour.

\section*{Acknowledgments} 
This research was partly financed by grant DEC-2011/03/B/ST9/03459 from the Polish 
National Science Centre and STFC grant ST/L00075X/1. P.Z. thanks the Kavli Institute for 
Particle Astrophysics and Cosmology for hospitality. A.I. acknowledges support from the 
Netherlands Organization for Scientific Research (NWO) Veni Fellowship, grant number 639.041.437.
We thank Bob Wagoner and Greg Madejski for discussions.

{}

\bsp


\begin{thebibliography}{}

 \bibitem[]{}
 Axelsson M., Done, C., 2016, MNRAS, 458, 1778
 \bibitem[]{}
 Axelsson M., Done, C., Hjalmarsdotter L., 2014, MNRAS, 438, 657
 \bibitem[]{}
 Bardeen J. M., Petterson J. A., 1975, ApJ, 195, L65
 \bibitem[]{}
 Beloborodov A. M., 1999a in High Energy Processes in Accreting Black Holes, ASP Conference Series 161, 
ed. J. Poutanen \& R. Svensson, p.295 (arXiv:astro-ph/9901108)
 \bibitem[]{}
 Beloborodov A. M., 1999b, ApJ, 510, L123 
\bibitem[]{}
  Churazov E., Gilfanov M.,  Revnivtsev M., 2001, MNRAS, 321, 759
\bibitem[]{}
  Done C., Gierli\'{n}ski M.,  Kubota A., 2007, A\&AR, 15, 1
\bibitem[]{}
  Fragile P. C., Mathews G. J., Wilson J. R., 2001, ApJ, 553, 955
\bibitem[]{}
  Fragile P. C., Blaes B. M., Anninos P., Salmonson J. D., 2007, ApJ, 668, 417
\bibitem[]{}
  G\'{o}recki A., Wilczewski W., 1984, AcA, 34, 141
\bibitem[]{}
  Ingram  A., Done C.,  2011, MNRAS, 415, 2323
\bibitem[]{}
  Ingram  A., Done C.,  2012a, MNRAS, 419, 2369
\bibitem[]{}
  Ingram  A., Done C.,  2012b, MNRAS, 427, 934
\bibitem[]{}
  Ingram  A., Done C., Fragile P. C., 2009, MNRAS, 397, L101
\bibitem[]{}
  Ingram  A., Maccarone T. J., Poutanen J., Krawczynski H., 2015, ApJ, 807, 53
\bibitem[]{}
  Ingram A., van der Klis M., Middleton M., Done c., Altamirano D., Heil L., 
  Uttley P., Axelsson M., 2016a, MNRAS, 461, 1967
\bibitem[]{}
  Ingram A., van der Klis M., Middleton M., Altamirano D., Uttley P., 2016b, MNRAS, 
in press (arXiv:1610.00948 [astro-ph.HE])
\bibitem[]{}
   Janiuk A., Czerny B., \.{Z}ycki P. T., 2000, MNRAS, 318, 180 
\bibitem[]{}
  Pottschmidt K. et al., 2003, A\&A, 407, 1039
\bibitem[]{}
  Pozdnyakov L. A., Sobol I. M., Syunyaev R. A., 1983, ASPRv, 2, 189
\bibitem[]{}
  Remillard R. A., McClintock J. E., 2006, A\&AS, 38, 903
 \bibitem[]{}
   Sobolewska M., \.{Z}ycki P. T., 2006, MNRAS, 370, 405
\bibitem[]{}
   Stella L., Vietri M., 1998, ApJ, 492, L59
\bibitem[]{}
  Stern B. E., Poutanen J., Svensson R., Sikora M., Begelman M. C., 1995, ApJ, 449, L13
\bibitem[]{}
 Veledina A., Poutanen J, 2015, MNRAS, 448, 939
\bibitem[]{}
 Veledina A., Poutanen J, Ingram A., 2013, ApJ, 778, 165 
\bibitem[]{}
  \.{Z}ycki P. T.,  Sobolewska M., 2005, MNRAS, 364, 891

\label{lastpage}

\end{thebibliography}
\end{document}